\documentclass[journal]{IEEEtran}
\usepackage{cite}

\ifCLASSINFOpdf
  \usepackage[pdftex]{graphicx}
  \DeclareGraphicsExtensions{.pdf,.jpeg,.jpg,.png}
\else
\fi
\ifCLASSOPTIONcompsoc
  \usepackage[caption=false,font=normalsize,labelfont=sf,textfont=sf]{subfig}
\else
  \usepackage[caption=false,font=footnotesize,farskip=0em]{subfig}
\fi
\usepackage{url}

\usepackage{multirow}

\usepackage{siunitx}
\DeclareSIUnit\fahrenheit{\SIUnitSymbolDegree F}
\DeclareSIUnit\mile{mi}
\DeclareSIUnit\foot{ft}
\DeclareSIUnit\voltampere{VA} 
\DeclareSIUnit\perunit{pu} 
\DeclareSIUnit\year{y}
\DeclareSIUnit\var{var} 

\usepackage[textsize = tiny,textwidth=7mm,disable]{todonotes}
\newcommand{\change}[1]{\textcolor{black}{#1}}

\begin{document}
%
\title{Physical Effects of Distributed PV Generation on California's Distribution System}

%
\author{Michael~A.~Cohen
        and~Duncan~S.~Callaway
\thanks{The authors are with the Energy and Resources Group, University of California, Berkeley,
CA, 94720-3050 USA e-mail: {macohen,dcal}@berkeley.edu}
\thanks{This work was supported by the California Solar Initiative RD\&D program and Robert Bosch LLC through its Bosch Energy Research Network program.}
}

%



\maketitle

\begin{abstract}
Deployment of high-penetration photovoltaic (PV) power is expected to have a range of effects -- both positive and negative -- on the distribution grid. The magnitude of these effects may vary greatly depending upon feeder topology, climate, PV penetration level, and other factors. In this paper we present a simulation study of eight representative distribution feeders in three California climates at PV penetration levels up to 100\%, supported by a unique database of distributed PV generation data that enables us to capture the impact of PV variability on feeder voltage and voltage regulating equipment. When comparing the influence of feeder location (i.e. climate) versus feeder type on outcomes, we find that location more strongly influences the incidence of reverse power flow, reductions in peak loading and the presence of voltage excursions.  On the other hand, we find that feeder characteristics more strongly influence the magnitude of loss reduction and changes in voltage regulator operations.  We find that secondary distribution transformer aging is negligibly affected in almost all scenarios.  
\end{abstract}


%
\IEEEpeerreviewmaketitle

\section{Introduction}
As the deployment of distributed photovoltaics (PV) accelerates, researchers and power industry professionals have increasingly attended to the impacts -- both positive and negative -- that PV might have on the distribution system. Areas of concern include PV's effect on~\cite{katiraei2011solar}:

\begin{itemize}
\item System losses
\item Peak load (which impacts capacity investments)
\item Transformer aging
\item Voltage regulator mechanical wear
\item Power quality, particularly voltage magnitude
\item Reverse power flow and its effect on protection systems
\end{itemize}

Prior work in this area consists largely of case studies that use simulations to examine a selection of these issues in detail for a single feeder or a single climate, e.g.~\cite{quezada2006assessment, shugar1990photovoltaics, woyte2006voltage, thomson2007impact, navarro2013monte, widen2010impacts}.  Results in these papers range from finding that distributed PV can cause resistive losses to \textit{increase} at relatively low penetrations to finding that resistive losses continue to decline up to very high penetrations.  Of those papers that examine the impact of PV on voltage excursions, results range from very positive (i.e. acceptable voltages at all penetration levels~\cite{widen2010impacts}) to negative (i.e. unacceptable voltages at high penetration levels~\cite{navarro2013monte}).  

However, because distribution systems are highly heterogeneous in terms of topology, climate and loads served, it can be difficult to draw useful generalizations from these case studies.  We are aware of only two existing studies that examine a diversity of climates and feeder architectures~\cite{paatero2007effects, hoke2013steady}.  In both cases, however, the simulations are driven with hourly solar irradiance data from a single location for each feeder.  Therefore these studies cannot provide insight into how cloud transients and geographic diversity of distributed PV systems will influence distribution system operation.  

The aim of this paper is to evaluate some of distributed PV's impacts across a diversity of conditions and to inform policy makers and utility decision-makers regarding how extensive these impacts might be at penetrations that are rare today but could be prevalent in the future.  The key points of distinction from earlier studies are that \todo{1} \change{we run simulations with a much more realistic PV data set} and examine a larger number of impacts, climates and feeder types.  In addition to studying voltage excursions, resistive losses, incidence of reverse flow and impact on peak loading -- as have the aforementioned papers, to varying degrees -- we report on voltage regulator operation and loss of life in secondary transformers.  The PV data set comprises highly distributed production from residential and small commercial PV systems recorded over a full year at time intervals as small as one minute.  These data allow us study the impacts caused by PV variability on feeder voltage and operation of voltage regulation equipment.  By looking at all these factors together across different climates, feeder types and PV penetrations, we gain insight into what drives both negative and positive effects of distributed PV in distribution systems.   This article is based on a prior conference paper~\cite{cohen2013modeling}, and expands it by covering more climates, adding a detailed comparison of simulated load shapes to actual load shapes, and presenting new observations about the importance of geographic diversity.

Our central findings have to do with resistive losses and voltage regulation.  As one might expect, feeder type -- rather than location -- has the strongest influence on the total reduction in resistive losses.  We also find that percent peak load reduction and incidence of reverse power flow depend more on location (climate) than on feeder type.  However, the most severe voltage problems appear to be a function of location, rather than feeder type.  As we will describe, impacts on voltage regulators are small and can either increase or decrease relative to a no PV baseline, depending on feeder type (and independent of location). 

Though we investigate a very large range of impacts in this paper, we acknowledge that there are other impacts that are outside of our scope.  For example, we did not investigate the impact of the harmonic content of PV inverters on power quality and transformer aging.  We also limit our investigation of protection equipment impact to assessing the prevalence of reverse flow conditions.  Furthermore, though our simulations captured the effect of phase imbalances that might occur from random placement of single phase PV on a three phase network, we did not investigate scenarios where we deliberately loaded one phase with more or less PV than others.  These omissions and others are due to space, data and modeling limitations, and they merit further systematic investigation in future research.

\begin{table*}[!t]
\renewcommand{\arraystretch}{1.25}
\caption{Summary of Simulated Feeder Characteristics and Figure Legend}
\label{table:feeders}
\centering
\begin{tabular}{l l l r r r r r r r r r r}
\hline
&
\multirow{2}{*}{\parbox[b][0.37in]{0.5in}{\textbf{Name*}}} &
\multirow{2}{*}{\parbox[b][0.37in]{0.75in}{\textbf{Serves\cite{schneider2008modern}}}} &
\multirow{2}{*}{\parbox[b][0.37in]{0.52in}{\raggedleft\textbf{Nominal Peak Load (\si{\mega\watt})\cite{schneider2008modern}}}} &
\multirow{2}{*}{\parbox[b][0.37in]{0.38in}{\raggedleft\textbf{Dist. Transformers}}} &
\multirow{2}{*}{\parbox[b][0.37in]{0.5in}{\raggedleft\textbf{Residential Load$\dagger\kern-0.75em$}}} &
\multirow{2}{*}{\parbox[b][0.37in]{0.35in}{\raggedleft\textbf{Approx Length (\si{\kilo\meter})}}} &
\multicolumn{3}{l}{\parbox[t]{0.7in}{\raggedright\textbf{Baseline Peak Load (\si{\mega\watt})}}} &
\multicolumn{3}{l}{\parbox[t]{0.9in}{\raggedright\textbf{PV Profiles Selected for Use}}}\\
& & & & & & & \textbf{Berk.} & \textbf{L.A.} & \textbf{Sac.} & \textbf{Berk.} & \textbf{L.A.} & \textbf{Sac.}\\
\hline
\multirow{8}{*}{\parbox[b][1.25in]{0.1in}{\includegraphics[height=1.24in]{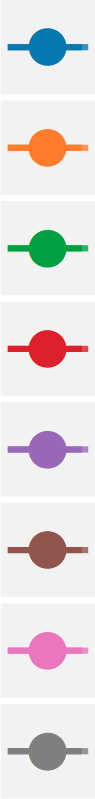}}}
& R1-12.47-1 & mod. suburban \& rural & 7.15 & 618 & 93\% & 5.5 & 5.56 & 5.38 & 7.59 & 21 & 38 & 26\\
& R1-12.47-2 & mod. suburban \& lt. rural & 2.83 & 264 & 84\% & 10.3 & 2.00 & 2.04 & 2.82 & 30 & 30 & 30\\
& R1-12.47-3 & moderate urban & 1.35 & 22 & 13\% & 1.9 & 1.27 & 1.25 & 1.60 & 10 & 10 & 8\\
& R1-12.47-4 & heavy suburban & 5.30 & 50 & 57\% & 2.3 & 4.31 & 4.09 & 5.65 & 12 & 17 & 12\\
& R1-25.00-1 & light rural & 2.10 & 115 & 2\% & 52.5 & 2.35 & 2.23 & 3.00 & 28 & 23 & 30\\
& R3-12.47-1 & heavy urban & 8.40 & 472 & 32\% & 4.0 & 6.64 & 6.30 & 8.70 & 20 & 31 & 25\\
& R3-12.47-2 & moderate urban & 4.30 & 62 & 0\% & 5.7 & 3.45 & 3.27 & 4.40 & 13 & 22 & 18\\
& R3-12.47-3 & heavy suburban & 7.80 & 1,733 & 84\% & 10.4 & 7.54 & 7.00 & 9.67 & 56 & 48 & 55\\
\hline
\multicolumn{13}{l}{* Climate region of origin is indicated by R1 (temperate west coast) or R3 (arid southwest). Nominal voltage is designated by 12.47 or 25.00 (kV).}\\
\multicolumn{13}{l}{$\dagger$ Approximate percentage of peak load that is residential, calculated from planning loads on the PNNL taxonomy feeders.}\\
\multicolumn{13}{l}{In figures, shape indicates Berkeley (\includegraphics[height=0.6em]{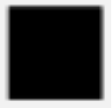}), Los Angeles (\includegraphics[height=0.75em]{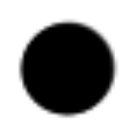}) and Sacramento (\includegraphics[height=0.75em]{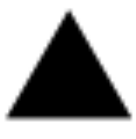}) results. Black symbols with dashed lines show means for each location.}\\
\end{tabular}
\end{table*}

\section{The Distribution Feeder Models}
\label{sec:models}
We used \mbox{GridLAB-D} version 2.3 (with the forward-backward sweep power flow solver) to model distribution circuits due to its integration of power flow analysis and time-varying load models, availability of representative feeder models, and open-source license.  \todo{2}\change{We used GridLAB-D's detailed load modeling capabilities for HVAC equipment (responsive to solar irradiance, outside air temperature and scheduled operation), residential water heating and pool pumps and commercial building lighting.  All remaining load at each building follows unique aggregated patterns that reflect variable occupancy and equipment scheduling.  Loads are modeled with detailed assumptions about power factor (see Section~\ref{sec:loads}) and ZIP load parameters; see~\cite{GLD_houseguide} for additional detail. In this section we describe our preparation of the models and supporting data.  }

\subsection{Feeder Topologies}
Pacific Northwest National Lab (PNNL) has compiled a set of representative ``taxonomy'' feeders drawn from utilities throughout the United States\cite{schneider2008modern}.  As summarized in \mbox{Table~\ref{table:feeders}}, the feeders vary along a number of important dimensions such as loads served (urban vs. rural), peak loading, and physical length. The feeders are organized by climate region. For this work, we selected the eight feeders originating from region~1 (temperate west coast) and region~3 (desert southwest) as these climates dominate California.

\subsection{Locations and Timeframe}
We simulated each of the eight feeders in three locations -- Berkeley, Los Angeles and Sacramento -- during the 366 days between September 25, 2011 and September 24, 2012, inclusive. We chose these locations and time span due to the availability of high-resolution PV generation and weather data. See Sections~\ref{sec:pv_data} to~\ref{sec:geographic} for more on this data and feeder placement. Note that the California peak demand during the selected year was fairly typical relative to the past decade, with a peak load of \SI{46846}{\mega\watt} in 2012 versus a high of \SI{50270}{\mega\watt} in 2006~\cite{caiso2013peak}. This means that the simulations do not include extreme conditions that may affect PV's overall value in important ways in the long run.

\subsection{Feeder Loads and Power Factors}
\label{sec:loads}
Because the taxonomy feeders specify only static planning (i.e. peak) loads, PNNL provides a script to populate the feeders with time-varying residential and commercial loads\cite{pnnl2012population}. Details of the loading process are discussed in detail in Sections 2.2-2.4 of~\cite{schneider2010evaluation}; we limit the discussion here to a few points of relevance.

The PNNL method models end-use loads with ``house'' objects that have a weather-dependent HVAC component and schedules for other types of loads such as appliances. The schedules for each house are scaled and time-shifted to provide heterogeneity among loads. Commercial loads are modeled as groups of ``houses'' with a different set of load schedules corresponding to commercial activities. 

The PNNL script applies a different distribution of load types depending on the climate region selected; e.g. air conditioning is more common in region~3 than in region~1. In this study, we applied region~3 loads to Los Angeles and Sacramento simulations and used region~1 loads in Berkeley, in keeping with the actual climate zone location of these cities.

Referring to the literature\cite{schneider2010evaluation, shoults2012power, bravo2012dynamic}, we adjusted the script-default load power factors as summarized in Table~\ref{table:pf}. We also reduced a capacitor bank on one feeder \mbox{(R1-25.00-1)} from \SI{150}{\kilo\var\per phase} to \SI{50}{\kilo\var\per phase} after noticing that it was overcompensating for reactive power, possibly because it is a rural feeder and is meant to handle more pumping load.

\begin{table}[!t]
\renewcommand{\arraystretch}{1.25}
\caption{Power Factors by Load Type}
\label{table:pf}
\centering
\begin{tabular}{l r | l r | l r}
\hline
\textbf{HVAC} & & \textbf{Residential} & & \textbf{Commercial} &\\
\hline
Base HVAC & 0.97 & Water heater & 1.0 & Int. lights* & 0.90\\
Fans & 0.96 & Pool pump* & 0.87 & Ext. lights* & 0.95\\
Motor losses & 0.125 & Other res.* & 0.95 & Plug loads* & 0.95 \\
                                          & & & & Street lights & 1.0\\
\hline
\multicolumn{6}{l}{* Power factor was changed from the PNNL default value of 1.0.}\\
\end{tabular}
\end{table}

\subsection{PV Generation Data}
\label{sec:pv_data}

The PV integrator SolarCity provided us with a database of instantaneous power at each inverter they monitor (roughly 7,000 systems, mostly in California) under the terms of a non-disclosure agreement.  All the inverters are single phase and provide data on the quarter hour; \todo{3}\change{for this project SolarCity also sampled a number of inverters at the fastest available time step of one minute.}

We performed data quality filtering to ensure we used only complete and credible profiles in the models. 
To address remaining missing readings in the selected profiles, we chose a very complete profile (with at least 365.8 days of non-zero readings between 8:00 and 16:00) from near the center of each location. We used readings from these ``filler'' profiles to fill gaps longer than one hour in other profiles from that location, scaling the filler readings by the ratio of the two profiles' rated capacity. Any shorter gaps we allowed to be handled internally by \mbox{GridLAB-D}, which uses the last-seen generation value until the model clock reaches the timestamp of the next reading.

\begin{table*}[!t]
\renewcommand{\arraystretch}{1.25}
\caption{Location Characteristics}
\label{table:locations}
\centering
\begin{tabular}{l | r r r | r r r | r | r}
\hline
& \multicolumn{3}{l|}{\textbf{Temp (\si{\celsius})}} & \multicolumn{3}{l|}{\textbf{Temp (\si{\fahrenheit})}} & 
{\textbf{PV Profiles}} & \multirow{2}{*}{\parbox[b][0.25in]{1.2in}{\textbf{Max Distance of PV Site from Weather Station}}}
\\
\textbf{Location} & \textbf{Low} & \textbf{Mean} & \textbf{High} & \textbf{Low} & \textbf{Mean} & \textbf{High} & 
\textbf{Used}\\
\hline
Berkeley & 0 & 13 & 35 & 32 & 56 & 94 &  97 & \SI{39}{\kilo\meter} (\SI{24}{\mile})\\
Los Angeles & 4 & 17 & 34 & 39 & 62 & 94 &  99 & \SI{27}{\kilo\meter} (\SI{16}{\mile})\\
Sacramento & -4 & 16 & 43 & 25 & 61 & 109  & 101 & \SI{45}{\kilo\meter} (\SI{28}{\mile})\\
\hline
\end{tabular}
\end{table*}

\subsection{Weather Data}
\label{sec:weather_data}
Table~\ref{table:locations} summarizes the weather data we used in this study.  We obtained one-minute temperature, humidity, and solar irradiance data for Berkeley from Lawrence Berkeley National Laboratoray~\cite{fernandes2012personal} and for Los Angeles and Sacramento from SOLRMAP at Loyola Marymount University and Sacramento Municipal Utility District\cite{nrel2012midc}. The Los Angeles and Sacramento data, having been quality controlled at the source, appeared to be quite complete and reliable and was used with only minor reformatting. 

The Berkeley data required the following edits: We calculated direct solar irradiance from global and diffuse irradiance using the solar zenith angle. Also, when irradiance data were missing or zero during the daytime, if less than an hour of data were missing we interpolated between adjacent values (for a total of 30 hours). For longer gaps (totaling 37.4 days) we copied in data from nearby days with similar cloud conditions as measured at Oakland Airport, \SI{18}{\kilo\meter} (\SI{11}{\mile}) south~\cite{noaa2013quality}. We also filled sub-hourly gaps in temperature data (totaling 5.5 days) by interpolation and longer gaps (totaling 25.6 days) directly with hourly measurements from Oakland Airport.

The temperature, humidity and irradiance data determined HVAC load in \mbox{GridLAB-D} but were not used to simulate PV generation, which was instead extracted from the SolarCity database. By using generation data sources located not far from the weather stations we preserved some (if not all) of the correlation between air conditioning load and PV generation. Given that buildings have significant thermal mass (resulting in a lagged and smoothed response to weather) and our goal was to preserve broad correlations between PV output and building load, we believe that the necessary corrections to the Berkeley weather data are acceptable and do not substantially affect the results.

\subsection{Geographic Assignment of PV Profiles}
\label{sec:geographic}
We sought to attach PV profiles to \mbox{GridLAB-D} houses in a way that reflects the diversity of solar generation over the area of a distribution feeder. This geographic diversity is driven in part by variations in cloud cover, but also by differences in PV system orientation, technology and shading -- all of which are reflected in the SolarCity data set.

The \mbox{GridLAB-D} taxonomy feeders are anonymized and therefore we do not know their physical layout. However, the models do contain electrical connectivity for all components and lengths for each overhead and underground line segment. We used this information and the graph layout utility Graphviz to create a geographic layout for each feeder subject to these constraints. These layouts are available online\cite{cohen2013taxonomy}.

We then used ArcGIS to superimpose the resulting feeder layouts on the SolarCity profile sources. We manually placed the feeders in locations with high densities of generation profiles to capture as much spatial diversity as possible. We then ran a ``nearest neighbor'' query to assign each distribution transformer to the closest SolarCity profile with acceptable data quality. As Table~\ref{table:locations} shows, at each location roughly 100 profiles were used (that is, matched with a transformer) with at least one feeder. Table~\ref{table:feeders} breaks down the number of profiles used in each individual scenario.

\subsection{Penetration Levels and PV Placement}
\label{sec:penetration}

For each \mbox{GridLAB-D} run, we populated only a portion of the houses with PV, to test various levels of penetration. To define ``penetration'' we first needed to establish a baseline loading for each feeder. To this end, we executed a baseline run for each feeder (with no PV) in each location and recorded its peak load. We then defined penetration as:
$$\mbox{PV penetration} = \frac{\sum \mbox{(PV system ratings)}}{\mbox{Peak feeder load from baseline run}}$$

We tested PV penetration levels of 0\%, 7.5\% 15\%, 30\%, 50\%, 75\% and 100\%.  We placed PV randomly across the available house models and used the same random number seed for all scenarios.  Using the same seed ensured that PV was placed at houses in the same order for each climate (Berkeley, Sacramento, Los Angeles), and that all systems populated in lower penetration runs were also populated in higher penetration runs.  This allowed us to make comparisons across climates and penetration levels.  We modeled the PV as a unity power factor ``negative load''. Each house's PV generation followed the time-varying load profile associated with its distribution transformer (as described in Section~\ref{sec:geographic}), scaled to an appropriate size for the building as described in Section~\ref{sec:pv_scaling}. Because GridLAB-D simulates three phase power flow and we randomly assigned PV systems to single phase points in the system, we are naturally capturing any phase imbalances that would occur from distributed PV in the specific case of random placement. To the extent these imbalances influence voltage magnitudes, they will influence our results in Sections~\ref{sec:regulators} and \ref{sec:voltage_quality}. 

All penetration levels should be treated as approximate for two reasons. First, our denominator for penetration was the baseline peak load during the test year, rather than the long-run feeder peak load which would typically be used in situations where more data was available. Second, due to transformer scaling (see Section~\ref{sec:xfmr_scaling}) and other minor adjustments, the peak loads from the final 0\% penetration runs differ slightly from the peak loads of our baseline runs. In general this difference is small, with the 0\% penetration runs having peak load ranging between 3.9\% lower and 2.9\% higher than the baseline runs. However, in one scenario (\mbox{R1-12.47-3, Berk.}) the final peak load was 8.0\% lower than the baseline peak load. So in this worst case scenario the nominal 100\% penetration might more accurately be read as a 108.7\% penetration.

\subsection{PV Generation Profile Scaling}
\label{sec:pv_scaling}

All of the selected PV generation profiles appear to be residential-scale, with system ratings ranging from \SI{1.68}{\kilo\watt} to \SI{13.16}{\kilo\watt}. To establish a reasonable installation capacity for each building, we first used the following formula from PNNL's load population script\cite{pnnl2012population}:
$$\mbox{building PV rating estimate} = A \times 0.2 \times 92.902$$

\noindent where $A$ is the floor area of the building in square feet, 0.2 is a rough estimate of the rated efficiency of the installations, and \SI[per-mode=symbol]{92.902}{\watt\per\square\foot} is the ``standard test conditions'' insolation.

We scaled up all commercial PV generation profiles so that their ratings matched this rating estimate. For residential installations, we scaled down the generation profile if its rating was \textit{higher} than the rating estimate for the house.  We did not scale up residential profiles with ratings smaller than the rating estimate since it is common for residential installations not to occupy the entire roof space.  We note that we did not simulate the effect of even larger standalone ``utility scale'' (multi-MW) PV systems.  Had we done so, we expect that voltage and reverse flow problems would be more severe than those we present in Sections~\ref{sec:voltage_quality} and~\ref{sec:backflow}.

\subsection{Transformer Scaling}
\label{sec:xfmr_scaling}
Transformer aging is one of our outcomes of interest, and it depends not on absolute loading of the transformer but loading relative to the transformer's rating \cite{ieee1996guide}. While the simulated loads are roughly scaled to the planning load value listed at each transformer in the taxonomy feeders, the loads may be somewhat larger or smaller than the planning loads due, for instance, to our use of different weather data at the three locations. This means that, unless corrected, some transformers would be sized inappropriately for the loads attached to them.

To address this issue, we assembled a ``menu'' of distribution transformers in standard \si{\kilo\voltampere} sizes based on the units present in the taxonomy feeders and manufacturers' data\cite{ge1972distribution,abb2001distribution}. We then replaced each transformer with the smallest transformer from the menu with a rating greater than the observed peak apparent power for that transformer from the baseline run. This is a conservative size estimate for distribution transformers given that in practice many carry power over their ratings during peak periods\cite{ieee1996guide}.

Note that to some extent the concern about transformer sizing also applies to conductor sizing; some taxonomy feeder line conductors may not be sized appropriately for the simulated loads.  Because conductor sizing was not a focus of this work, we did not undertake to resize the conductors in the way we did the transformers, and indeed when we run \mbox{GridLAB-D} we occasionally observe warnings that conductors are modestly overloaded. This may slightly distort the absolute results for line losses. To address this we instead report the percent change in losses between penetration scenarios. The percent change should not be affected significantly by conductor size since line resistance is a linear scaling factor on line losses and all penetration levels use the same conductors.

\subsection{GridLAB-D Configuration}
All of the taxonomy feeders have an on-load tap changer (LTC) at the substation, and two of them feature additional line voltage regulators. During the baseline runs, we observed that the upper bound of the LTC and regulator deadbands were set at approximately \SI{1.05}{\perunit}, right at the edge of ANSI standards for end-use voltages. This contributed to a significant number of voltage violations due to time lag in regulator response when voltages rose outside the deadband. We therefore lowered the top of the LTC and regulator deadbands to \SI{1.04}{\perunit} (maintaining the bandwidth) for our production model runs. \todo{4} \change{The controller deadband is $\pm$\SI{0.08}{\perunit} on all voltage regulators and LTCs.}

\todo{5} \change{GridLAB-D runs with an adaptive time step, meaning that it runs the power flow solver only when an input to the model (such as weather or PV production) changes or a simulated element within the model is expected to change (for example a building model).  As described above, the PV data we used were sampled at most once per minute, and we used 1 minute resolution weather data.  Because the data inputs change no more than once per minute, simulated voltage regulating equipment will not change position more than once per minute.  Therefore to contain run time we set the minimum simulation time step to 1 minute.  We note that in practice voltage regulating equipment may have shorter delay times (e.g. 30 seconds); as we will address in the results section (Sec.~\ref{sec:regulators}), we believe that limiting the time step to 1 minute does not significantly affect the results.  }

\todo{6} \change{See~\cite{GLD_powerflowguide} for additional detail about GridLAB-D configurations.}

\section{Results}

Our results must be interpreted with several important caveats in mind. First, the simulation covers one particular year that was chosen primarily for PV data availability. It may not include extreme weather or other events that would drive true system peaks in the long term. Second, though the \mbox{GridLAB-D} load models are physically-based and the taxonomy feeders are based on real feeders, we did not model the actual feeders and loads in the study locations. Third, the prototypical feeders are ``typical'', meaning they do not have special problems such as poor voltage regulation or capacity constraints that would require special attention when integrating PV.  Finally, because of the transformer and conductor sizing issues discussed in Section~\ref{sec:xfmr_scaling} and later in Section~\ref{sec:transformer_aging}, we do not consider thermal overloading of transformers and lines; this prevents us from being able to make specific claims about the hosting capacity of each network.

The base quantity for all normalized results is the value of the metric in question for the feeder at 0\% penetration.

\subsection{System Losses}

\begin{figure}[!t]
\centering%
\subfloat[Normalized system losses.]{\includegraphics[width=1.7in]{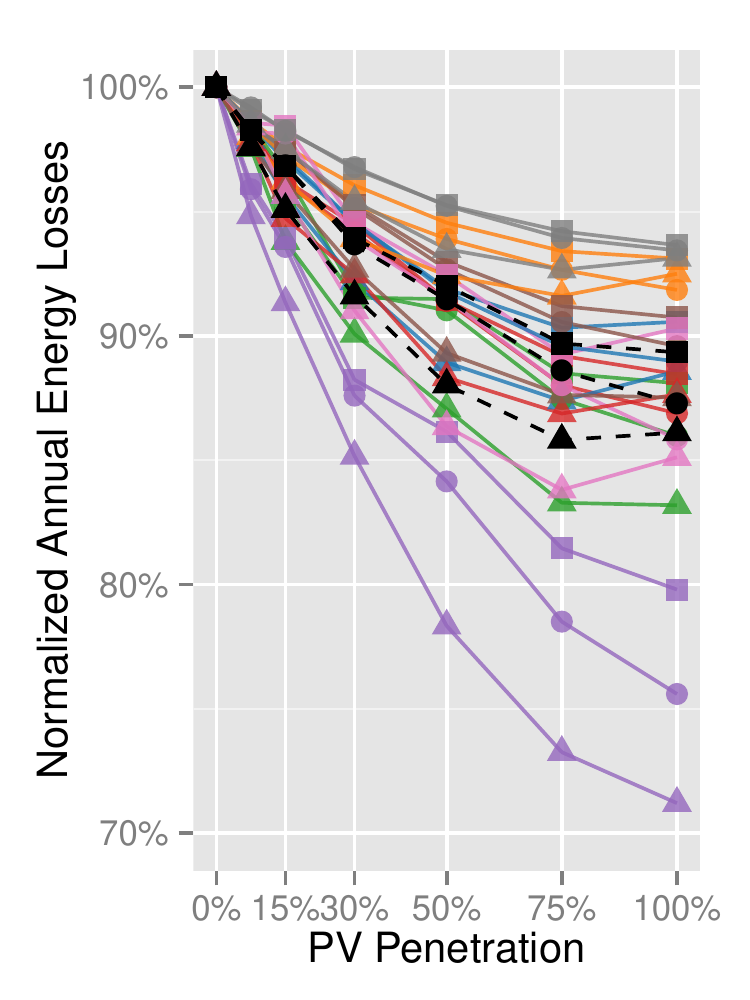}%
\label{fig:losses}}
\hfil
\subfloat[Losses as a percentage of load energy supplied by the grid.]{\includegraphics[width=1.7in]{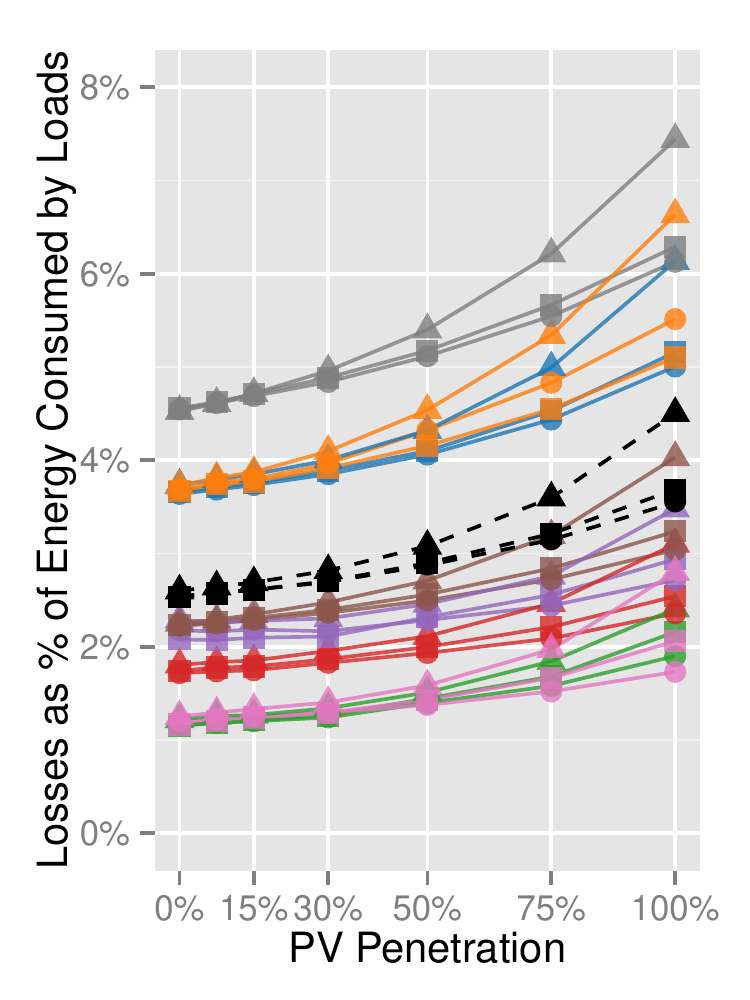}%
\label{fig:loss_vs_load}}
\caption{Left: Normalized system losses. See Table~\ref{table:feeders} for key.}
\label{fig:losses_parent}
\end{figure}

We measured instantaneous system losses (including transformer and line losses) every fifteen minutes. As shown in Figure~\ref{fig:losses}, we found that increasing PV penetration decreased system losses, with diminishing effects at high penertations. The impact of PV on losses was similar across the three locations, but varied considerably by topology, with losses reduced by anywhere from 7\% (\mbox{R3-12.47-3}) to 28\% (\mbox{R1-25.00-1}) at 100\% penetration. In particular, feeders with higher nominal peak loads (see Table~\ref{table:feeders}) tended to have less loss reduction with increasing PV, though this trend was not universal.  We also found, unsurprisingly, that the feeder that experienced the largest reduction in percent losses was also the longest.   For reference, the total annual losses in the 0\% penetration case for each feeder are:   810 MWh (R1-1247-1), 310 MWh (R1-1247-2), 40 MWh (R1-1247-3), 290 MWh (R1-1247-4), 150 MWh (R1-2500-1), 540 MWh (R3-1247-1), 130 MWh (R3-1247-2), 1440 MWh (R3-1247-3).  Note that to produce these statistics we averaged across each of the three locations for each feeder and we rounded to the nearest 10 MWh.  The actual values for each feeder varied on the order of $\pm$ 5\% across locations.

We attribute the reduced marginal effect of PV at high penetrations to the fact that losses are proportional to current squared; the more PV reduces power (and thus current) flow on the lines, the less effect further reductions will have on losses. For some feeders (mainly in Sacramento) losses \emph{increased} as penetration rose from 75\% to 100\%, presumably because the losses associated with high ``backflow'' currents at certain times began to exceed the losses ``saved'' at other times when net current flow was lower.

Other studies have found that resistive losses increase with penetration~\cite{quezada2006assessment,widen2010impacts,navarro2013monte,thomson2007impact}.  However in contrast to other work, our finding is that on \textit{most} feeders we study, losses continue to decline up to 100\% penetration.  We note that in the feeder / location pairs here, location seems to determine whether or not losses begin to increase in the range of penetrations we examined, but that the total magnitude of losses is much more strongly influenced by the feeder type.  

Figure~\ref{fig:losses} shows that losses as a percentage of energy consumed by loads from the grid (i.e. as a percentage of utility wholesale power purchases) generally increase with PV penetration. This is likely because most of the load reduction happens off-peak, when system losses are lower than on-peak.

\subsection{Peak Loading}
\label{sec:peak_load}

We measured peak load as the maximum fifteen-minute rolling average of one-minute measurements at the substation. The extent to which PV reduces feeder peak load depends largely on the timing of the peaks. Clearly, peak load reduction will be greatest if peak load is coincident with peak PV production.  In California, however, load typically peaks later in the day than PV production, and therefore peak loads are reduced by only a fraction of the PV's rating.

\begin{figure}[!t]
\centering
\subfloat[Normalized peak loads.]{\includegraphics[width=1.7in]{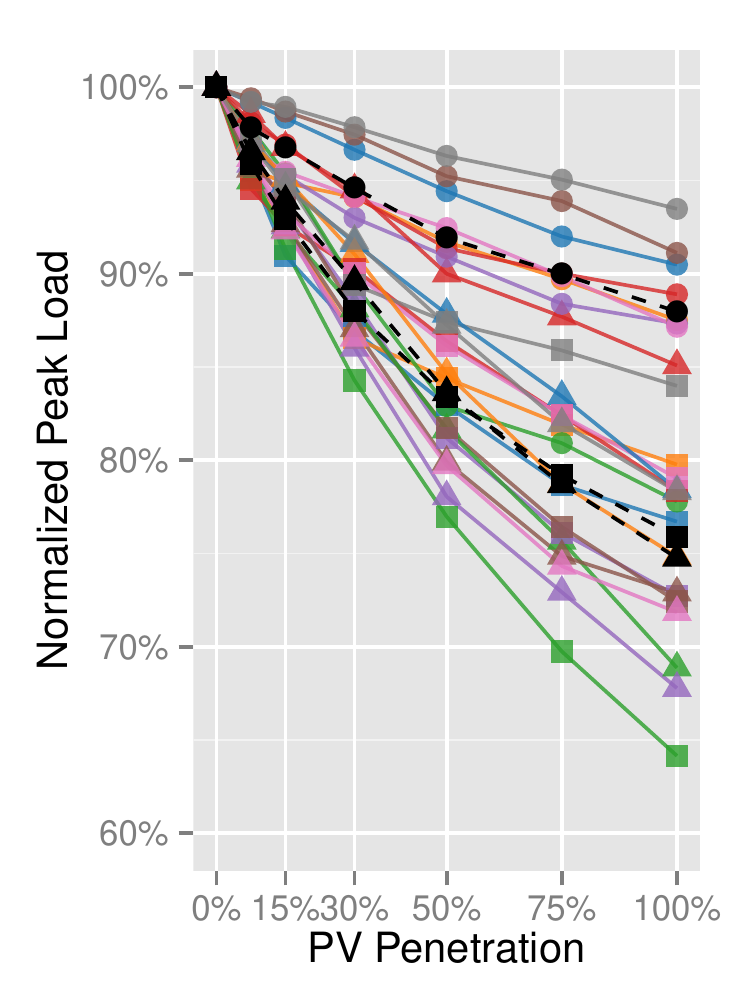}%
\label{fig:peak_load}}
\hfil
\subfloat[Ratio of peak load reduction to penetration]{\includegraphics[width=1.7in]{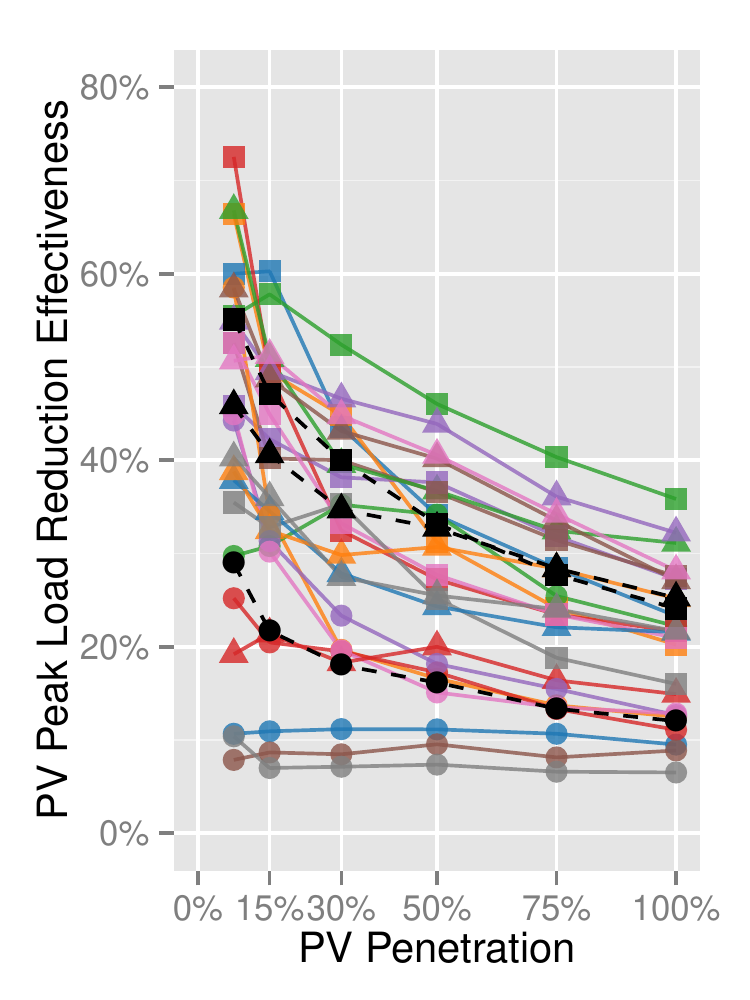}%
\label{fig:solar_effectiveness}}
\caption{Effect of PV on peak loads. See Table~\ref{table:feeders} for key.}
\label{fig:peak_loads}
\end{figure}

As shown in Figure~\ref{fig:peak_loads}, we observed that PV generally reduced peak loads by much less than the penetration percentage. \todo{7} \change{This is due to the fact that peak load occurs several hours after peak PV production.}   In contrast to system losses, location (i.e. climate) had a strong effect on the peak load reduction impact of PV, with Sacramento and Berkeley showing more significant reductions than Los Angeles. Figure~\ref{fig:peak_load} shows the normalized peak load as a function of PV penetration, whereas Figure~\ref{fig:solar_effectiveness} shows the peak reduction as a percentage of the solar penetration. Figure~\ref{fig:solar_effectiveness} illustrates that low penetrations of PV can be quite effective at reducing peak loads, although this is not true in all cases. Peak load reduction effectiveness diminishes as penetration increases because early increments of PV tend to reduce daytime peaks, causing the new peak to be in the evening when PV contributes less power.

\begin{figure}[!t]
\centering
\subfloat{\includegraphics[width=1.7in]{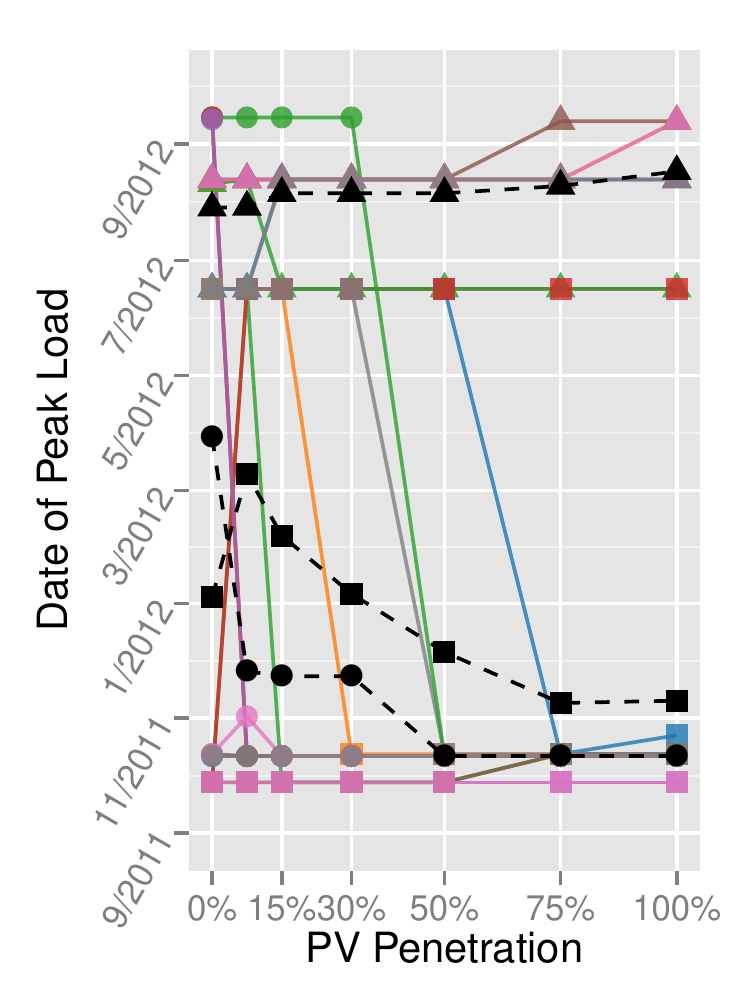}%
\label{fig:peak_date}}
\hfil
\subfloat{\includegraphics[width=1.7in]{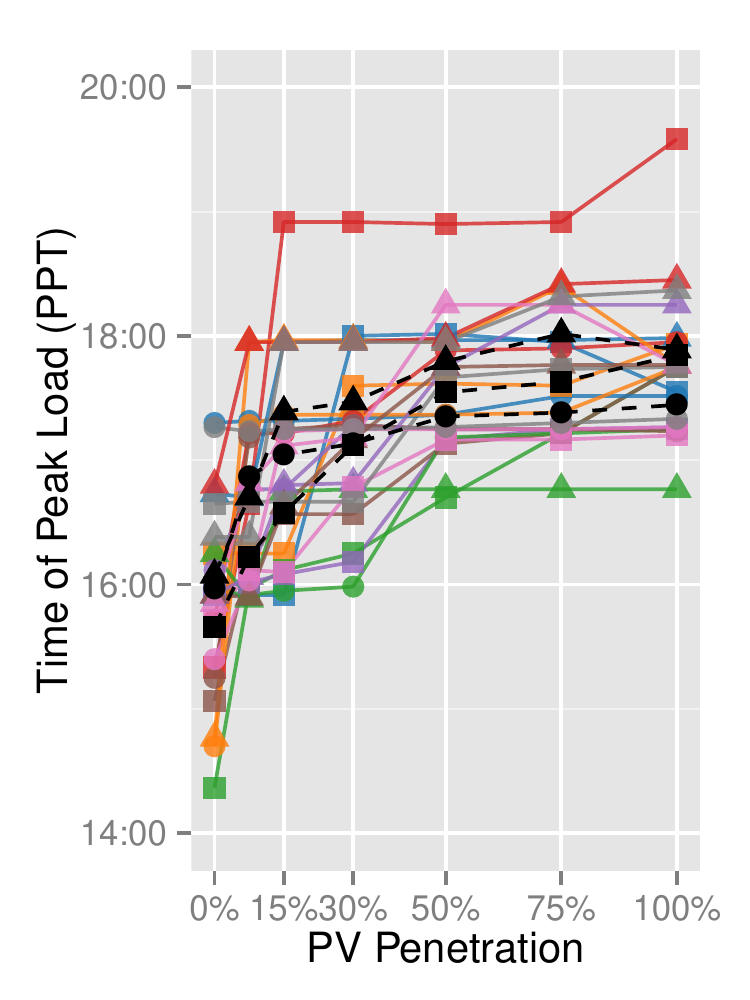}%
\label{fig:peak_time}}
\captionsetup{aboveskip=-5pt}
\caption{Date and time of peak loads. The time reported is the first minute of the peak fifteen-minute period.  See Table~\ref{table:feeders} for key.}
\label{fig:peak_date_time}
\end{figure}

Figure~\ref{fig:peak_date_time} illustrates trends in the timing of peaks as PV penetration increases. Without PV, peak loads arrived in August 2012 for most Sacramento feeders and half of the Los Angeles feeders, while Berkeley feeders generally peaked in fall 2011 or June 2012. Peak times were widely dispersed between 14:22 and 17:18. However, a 7.5\% penetration of PV was sufficient to eliminate August peaks for all but one Los Angeles feeder, shifting their peaks to the later afternoon during a relatively warm spell in October 2011. Berkeley peaks, while initially shifting towards the summer, were ultimately also moved to the fall by high penetrations of PV. Meanwhile the Sacramento peaks, driven by larger air conditioning loads, remained in the summer at all levels of penetration, although moving noticeably later in the afternoon. In all locations, peaks were moved later in the day as PV reduced daytime usage.

\begin{figure*}[!t]
\centering
\includegraphics[width=5in]{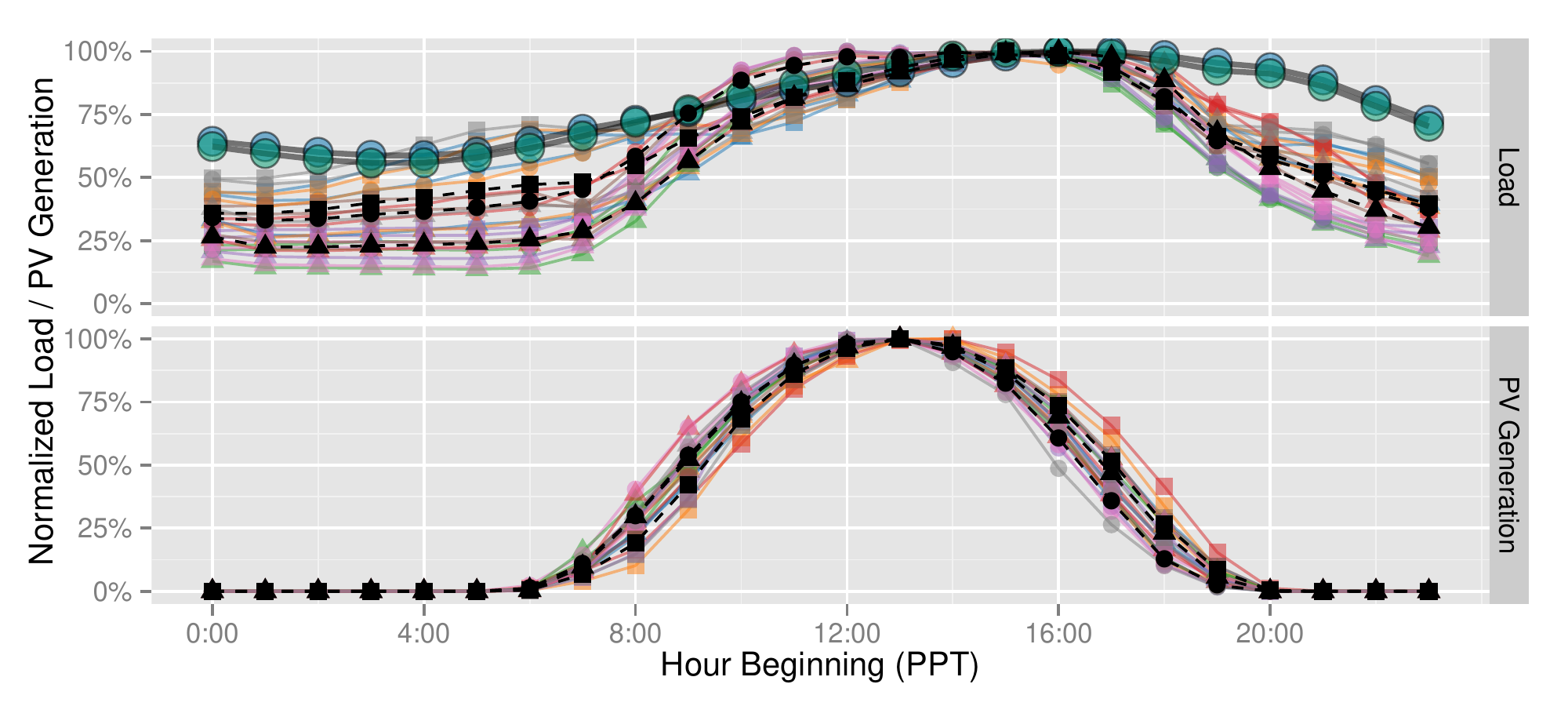}
\captionsetup{aboveskip=-5pt}
\caption{Normalized hourly load and PV generation profiles for August 13, 2012. Normalized PG\&E system load is shown by larger blue circles and CAISO load by larger green circles~\cite{caiso2013oasis}.}
\label{fig:load_profiles}
\end{figure*}

When interpreting the peak load reduction results, it is important to consider how well the simulated feeder load shapes align with feeder load shapes actually found in California. We do not have access to a large enough corpus of load shapes to do a rigorous analysis of this issue, but a high-level comparison will suffice to contextualize our findings. Figure~\ref{fig:load_profiles} shows the average hourly load and PV generation for each of the simulated feeders on August 13, 2012, which was the day CAISO recorded its peak demand for 2012~\cite{caiso2013peak}. It is also the peak demand day for five simulated Sacramento feeders, though not for any Los Angeles or Berkeley feeders. Each individual profile is normalized against the peak hour for that profile. As in the other figures, the locational means are straight averages of the eight normalized feeder simulations, i.e. the feeders are not weighted by their size or expected frequency of occurrence in the field. The load plot also shows normalized CAISO system load (larger green circles) and PG\&E system load (larger blue circles).

From this figure we can see that the simulated peaks match well with the PG\&E and CAISO peaks in the 15:00-16:00 range. However, the simulated feeders universally drop in demand more quickly than the CAISO system. Note from the bottom panel in Figure~\ref{fig:load_profiles} that PV production goes to zero after the simulated load drops, but before any significant drop in CAISO load.  This suggests the possibility that peak demand might be relatively unaffected by PV in the CAISO system, but strongly affected in our simulations.

This simple one-day comparison ignores several factors that are important when calculating annual peak demand reduction, such as load variation within each hour and the fact that PV often shifts the peak to a different day, rather than a different time on the same day. Also, the comparison to an overall system load profile greatly obscures the wide variation of individual feeder profiles that comprise it. For instance, SCADA data provided by PG\&E under the terms of a nondisclosure agreement indicates that on August 13, 2012 the most common hours for feeders to peak were 16:00 and 17:00, but each of these hours only accounted for about 16\% of feeders, with 37\% peaking earlier (including 10\% before noon) and 31\% later in the evening~\cite{carruthers2013personal}. Thus, it is likely that the simulated load shapes are a good match to some subset of California feeders and therefore the reported peak load reduction is achievable in some locations. However, the fact that the simulated feeder profiles are not a good match for the general system profile in the evening indicates that it would be optimistic to expect the simulated peak load reduction to occur universally across California.

\subsection{Transformer Aging}
\label{sec:transformer_aging}

\mbox{GridLAB-D} 2.3 implements the IEEE Standard C57.91 \mbox{Annex G}\cite{ieee1996guide} method for estimating transformer insulation aging under various loading conditions. \mbox{GridLAB-D} implements the method for single phase center tapped transformers only. This is the most common type of transformer on the taxonomy feeders, but one feeder (\mbox{R3-12.47-2}) did not have any so it was excluded from the aging analysis. In the \mbox{Annex G} model, a ``normal'' year of aging corresponds to the amount of insulation degradation expected if the transformer hot spot were at a constant \SI{110}{\celsius} throughout the year. A transformer that is often overloaded will age more than \SI{1}{\year} in a year, and thus may need to be taken out of service due to insulation degradation before its rated lifetime. On the other hand, one that is loaded below its rating will age less than \SI{1}{\year} per year, and will be unlikely to have its insulation fail prematurely.

In general, we observed minimal aging in all scenarios and penetration levels, with a mean equivalent aging of up to \SI{0.29}{\year} in one scenario (\mbox{R3-12.47-3}, Sac.) and all other scenarios having mean aging less than \SI{0.001}{\year}. We attribute this slow aging to the fact that the transformers were conservatively sized at or above their baseline peak load (see Section~\ref{sec:xfmr_scaling}). However, in \mbox{R3-12.47-3} (Sac.) at PV penetrations of 30\% and above we did observe a small number of transformers aging quite rapidly, up to \SI{166}{\year} during the simulated year (all other scenarios had maximum individual transformer aging less than \SI{0.38}{\year} per year). These few rapidly aging transformers are likely at a location where net PV generation is often higher than the load they were sized to handle, and in reality they would need to be upgraded to handle this backflow.

\subsection{Voltage Regulators}
\label{sec:regulators}

Tap-changing voltage regulator wear and tear is driven primarily by the number of tap changes the device must perform and the current that it handles during operation. In our simulations, tap changes at the substation LTC were on the order of 20 per day.  However the count was not affected by topology, climate or PV penetration, varying between 7,166 and 7,243 changes across all model runs over the year of simulation -- a difference of only 1\%. This small difference is because the models did not include a transmission impedance component, with the transmission voltage instead following a fixed schedule of values recorded from an actual substation in the U.S. Western Interconnection (WECC). The substation LTC operates to maintain voltage immediately downstream within the deadband despite fluctuations in the WECC schedule, and is insensitive to downstream changes in load. Due to the lack of a transmission model, our simulations do not provide reliable insight on LTC response to PV.

\begin{figure}[!t]
\centering
\subfloat[Tap changes.]{\includegraphics[width=1.7in]{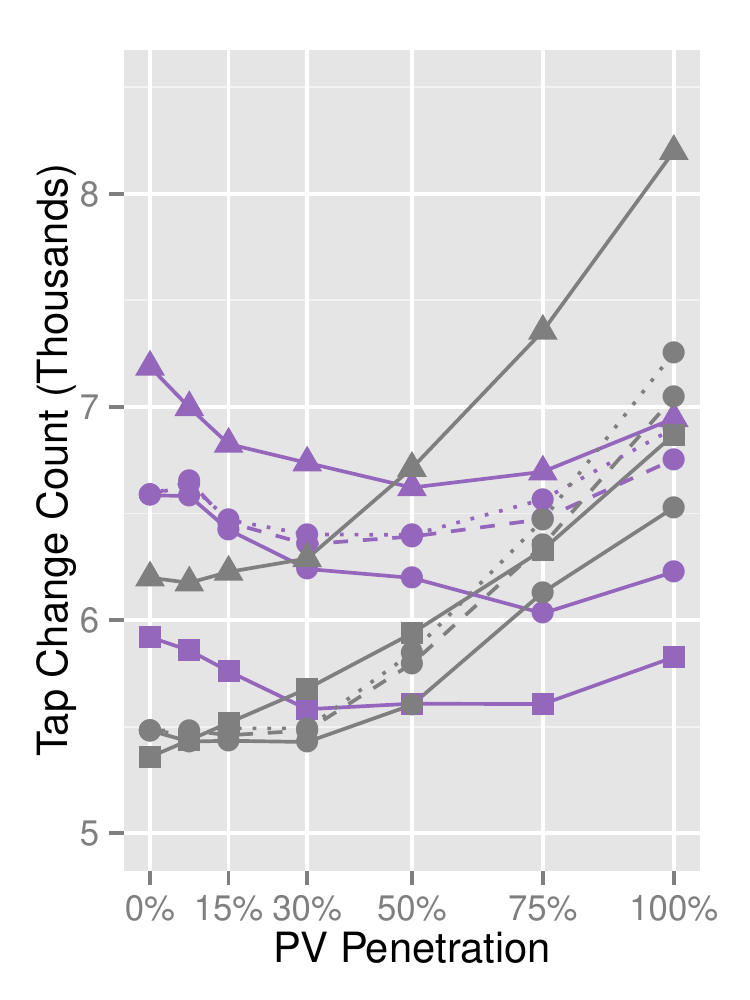}%
\label{fig:tap_changes}}
\hfil
\subfloat[Mean current duty.]{\includegraphics[width=1.7in]{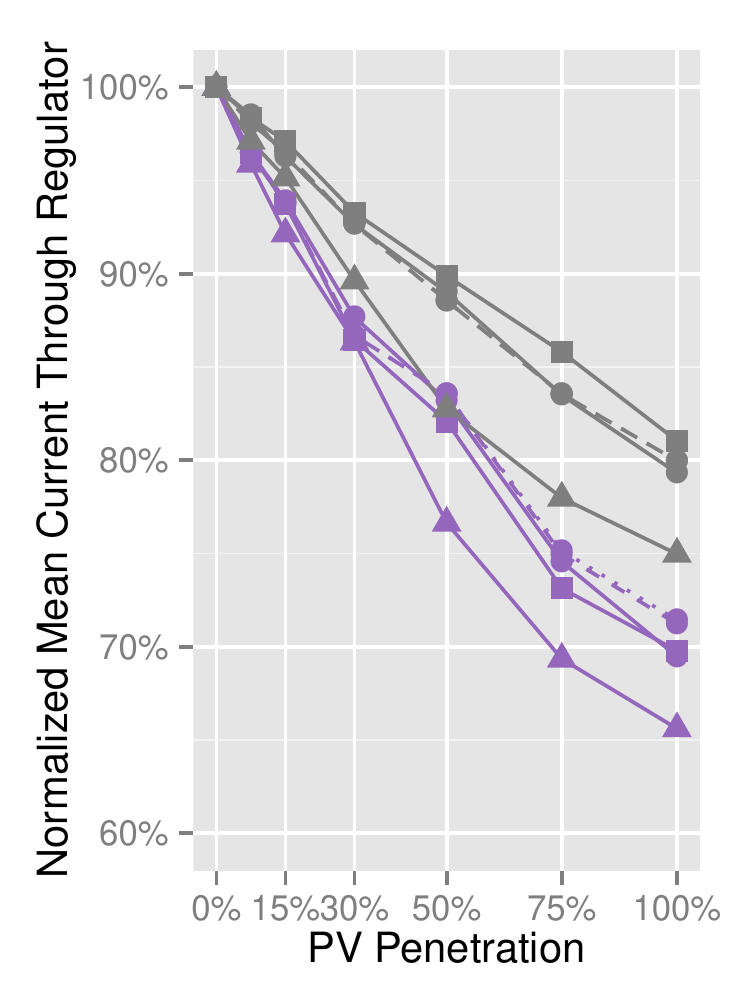}%
\label{fig:avg_current}}
\caption{Line voltage regulator activity across all three phases. See \mbox{Section~\ref{sec:regulators}} for discussion of broken lines.}
\label{fig:regulators}
\end{figure}

The two mid-feeder regulators in the simulation (at \mbox{R1-25.00-1} and \mbox{R3-12.47-3}) do have simulated impedances and varying loads both upstream and downstream and thus exhibit more variation. Figure~\ref{fig:tap_changes} shows that PV has little effect at \mbox{R3-12.47-3} until 50\% penetration, at which point tap changes begin rising noticeably. This result echoes other work\cite{mather2012quasi} and concerns from utilities that PV variability will increase regulator maintenance needs. \todo{8} \change{However, the total change in the number of control actions is relatively small.  We believe this is due to the fact that fast time scale variability in PV output is a relatively small amount of the total variability in PV output~\cite{lave2013wavelet}, and therefore the number of control actions is largely driven by the diurnal range of net load.    At low to moderate penetrations, the range of net demand has the tendency to \textit{decrease} as PV reduces peak demand but does not push mid-day demand below the night time minimum.  However at higher penetrations, the range of net demand grows as peak net demand is relatively unaffected (see Fig.~\ref{fig:peak_loads}) but mid-day net demand begins to drop below the night time minimum.   These results indicate that in some cases PV could in fact \textit{reduce} voltage regulator maintenance needs at intermediate penetrations.}

We examined two sensitivity scenarios to study the impact that the PV data had on the regulator results. To produce the dotted lines in Figure~\ref{fig:regulators} we used the single PV profile with the most one-minute data available (82\% of days) at all PV sites. The dashed line shows the same scenario with the one-minute data downsampled to fifteen-minute resolution; this intermediate scenario helps us to distinguish the effect of the one-minute data from the effect of eliminating geographic diversity. We limited the sensitivities to Los Angeles because this was our source of one-minute data. Figure~\ref{fig:tap_changes} suggests that geographic diversity reduces tap change frequency (because the solid lines which include geographic diversity fall well below their corresponding single-profile dotted and dashed lines) and that fifteen-minute PV data is a reasonable proxy for one-minute data when studying regulator behavior (because the dashed lines track their corresponding dotted lines closely).  

\todo{9} \change{It is possible that with PV data on even finer time scales (faster than once per minute) a different pattern of regulator activity would emerge.  However, we hypothesize that this is not the case for several reasons.  First, as we discussed in the previous paragraph, the total amount of regulator action appears to be driven by diurnal variability (a daily occurrence) rather than partly cloudy conditions.  Second, since regulators generally have a response lag on the order of \SI{30}{\second}, very brief fluctuations in PV are likely to result in voltage changes on the feeder rather than increased regulator activity.}

The effect of PV on regulator current duty was more consistent than the effect on tap changes, as illustrated by Figure~\ref{fig:avg_current}. With PV reducing the downstream load, current through the regulator declines steadily as penetration increases. This suggests that even in cases where PV increases a regulator's activity, its expected lifetime may stay the same or even increase because each tap change is less destructive under lighter current duty. Our sensitivity runs suggest that neither geographic diversity nor the use of one-minute resolution data has a substantial effect on regulator current duty.

\begin{figure}[!t]
\centering
\subfloat[Proportion of voltages outside ANSI standards.]{\includegraphics[width=1.7in]{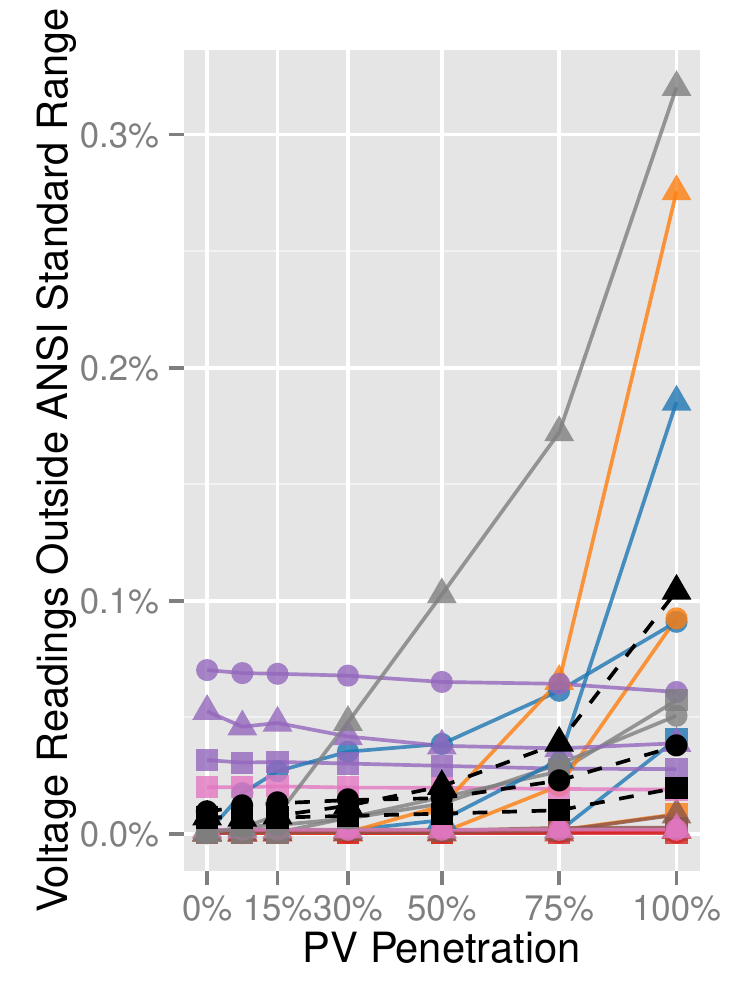}%
\label{fig:bad_voltages}}
\hfil
\subfloat[Annual minimum load.]
{\includegraphics[width=1.7in]{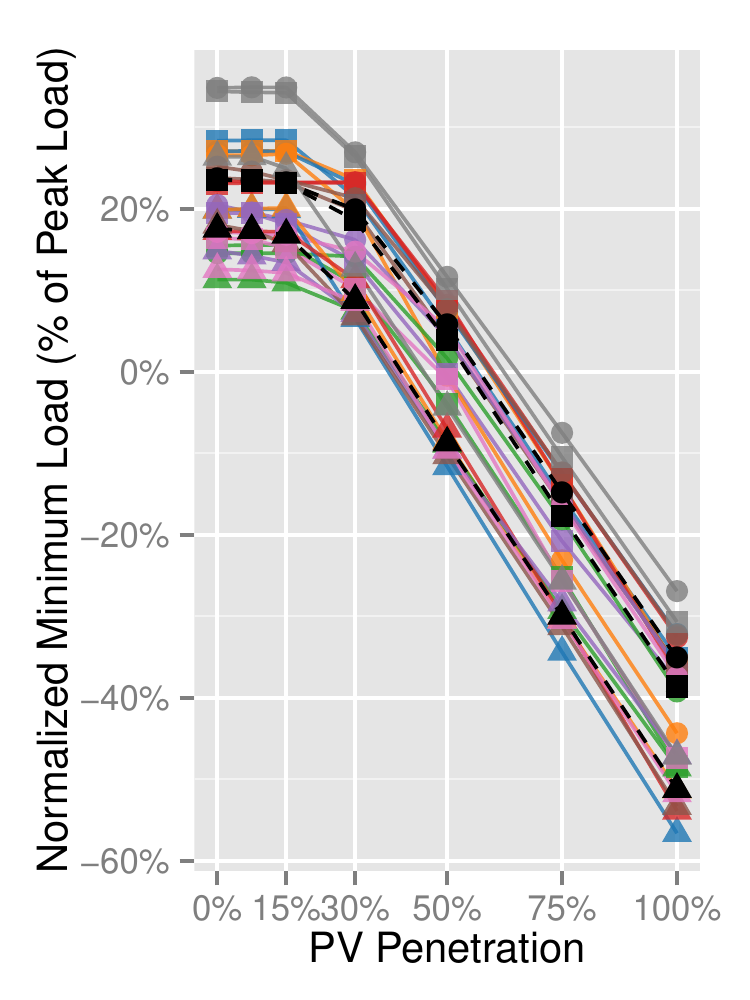}%
\label{fig:backflow}}
\caption{Voltage control and minimum load (representing the magnitude of reverse power flow). Many scenarios overlap near 0.0\%.}
\label{fig:voltage_and_backflow}
\end{figure}

\subsection{Voltage Quality}
\label{sec:voltage_quality}

We recorded voltage at all point-of-use meters at fifteen minute intervals and tabulated in Figure~\ref{fig:bad_voltages} the proportion of readings falling outside of the ANSI standard range of \SIrange[range-phrase = --]{0.95}{1.05}{\perunit}. In general, voltages appear to be well-controlled, with most runs having less than 0.002\% of readings out of range, and the worst case (\mbox{R3-12.47-3}, Sac.) having 0.32\% of readings out of range. \todo{10} \change{This is consistent with prior work suggesting that many feeders can support high penetrations of PV without voltage violations~\cite{hoke2013steady}, however it may be counter-intuitive that feeders designed for one-way power flow can host so much PV capacity without more negative voltage impacts.  There are several explanations for this.  First, the feeders we investigated had relatively good voltage control and voltage regulators rarely saturated; it is plausible that there are feeders in operation whose control is more likely to saturate.  Second, we did not model scenarios with PV heavily concentrated in part of a feeder -- this would exacerbate local reverse power flow and voltage rise.  Finally, though the maximum penetration we investigated is relatively high, penetrations could be on the order of 200\% if systems were sized to produce as much energy over the course of a year as each building consumes.   We expect that voltage excursions would be much more significant at those penetrations.}

In general, the voltage violations that did occur took place on rural and suburban feeders (see Table~\ref{table:feeders}) with violations being very rare on urban feeders at all penetration levels. Except at feeder \mbox{R1-25.00-1}, almost all out-of-range voltages observed were greater than \SI{1.05}{\perunit}. As expected these high-side excursions generally become more frequent as penetration increased and the power injection from PV raised some voltages locally. At \mbox{R1-25.00-1} the out of range voltages were predominantly less than \SI{0.95}{\perunit}, with a small amount greater than \SI{1.05}{\perunit}. Under these conditions, increasing PV penetration improved voltage quality on the feeder by boosting some local voltages that would otherwise be low. As noted in Section~\ref{sec:regulators}, it is possible that more brief voltage excursions would be observed with higher resolution PV generation data.

\subsection{Reverse Power Flow}
\label{sec:backflow}

\todo{11} \change{Figure~\ref{fig:backflow} shows the minimum load, as a fraction of peak demand, measured over the year of simulation on each feeder.  Negative values indicate that the feeder experiences reverse power flow conditions.  These results indicate that the amount of reverse power flow takes on a very large range across the feeders we investigated, and that Sacramento feeders experience the largest reverse power flow conditions.  This result is due to the fact that Sacramento loads have larger peak to mid-day demand ratios (due to air conditioning loads peaking in the late afternoon or early evening); PV penetration is defined by peak demand but reverse power flow depends PV production and \textit{mid-day} demand.}

We also investigated the incidence of negative real power flow (``backflow'') through the substation, which can be a proxy for protection issues and higher interconnection costs. At 50\% penetration, twelve of the 24 scenarios exhibited occasional backflow, up to 1\% of the time each. At 100\% penetration, all scenarios experienced backflow at least 4\% of the time. In general, backflow was more prevalent in Sacramento because PV penetration in Sacramento was measured against a higher peak air conditioning load. This led to a larger absolute quantity of PV generation in Sacramento but with similar low loads to Los Angeles and Berkeley on cooler days.

\subsection{Observations Regarding Geographic Diversity}
\label{sec:diversity}

We ran our sensitivity scenarios primarily to assess the effect of PV profile time resolution and geographic diversity on voltage regulator operation (see Section~\ref{sec:regulators}). However, these scenarios enable us to observe how other outcomes vary with the input data as well. These observations are necessarily tentative because the sensitivities were run for only two feeders (R1-25.00-1 and R3-12.47-3) in one location (Los Angeles).

First, we note that for all outcomes observed, differences between the single-profile one-minute input and that input downsampled to fifteen-minute resolution were minimal. This implies that fifteen-minute PV data is ``good enough'' for a reliable study of PV's effects on the distribution system.

Second, for two metrics we did observe changes in outcomes when switching from the full geographic diversity of profiles to the single profile for all PV installations. First, peak load reduction was larger with geographic diversity than without it. We attribute this to the fact that the diverse set of profiles includes west-facing installations that are more effective at reducing peak load. We also noticed substantially less backflow at high penetrations with geographic diversity. This is expected because with a single profile periods of high generation will be completely coincident, whereas with a diverse set of profiles they will be spread out somewhat -- by system orientation if not by cloud cover differences -- reducing the overall ``peakiness'' of PV generation and thus backflow. Taken together, these observations suggest that studies that do not account for the geographic diversity of PV -- even on a distribution feeder scale -- may underestimate some of its benefits and/or overstate its drawbacks.

\section{Concluding remarks}

We studied how distributed PV impacts distribution systems across a variety of feeder architectures and climates within California over a full year of operation.  In contrast to earlier studies, we ran simulations with real PV data (either 1-minute or 15-minute resolution), which allowed us to uniquely address issues of voltage regulation on the time scale of cloud transients.  In addition to studying voltage excursions, resistive losses, reverse flow and impact on peak loading -- as have researchers before us --  we examined voltage regulator operation and loss of life in secondary transformers.  We used unique PV data that captured the impacts of fast cloud transients, array shading and spatial diversity.

It is worth emphasizing that, while this paper is extensive in terms of its combination of geographic scope, number of feeder types and high resolution PV data, it is not an exhaustive assessment of all possible outcomes.  Overall, we observe that undesirable impacts are relatively small in most cases but large in some; we expect that a similar pattern of observations would hold across an even larger range of California scenarios than we consider in this paper. \todo{12} \change{However the taxonomy feeder models we used are representative of a very diverse system of infrastructure.  Within that diversity it is likely that there are feeders that would experience more severe impacts from distributed PV.  In this sense we regard our results to be representative of typical feeders -- but not an exhaustive representation of the possible range of impacts.  Moreover, we are unable to say whether our results would hold in areas with very different climates, or different histories of electric grid development, e.g. areas that have invested less in voltage regulation than the U.S.  The research community would benefit from similar analyses with additional feeder models in additional locations to generalize the results in this paper.  There is also a need for additional measurement and verification in real feeders to understand how well model results reflect reality in these circumstances. }

\todo{13} \change{We also note that we have not studied measures to mitigate the observed impacts.  For example, if one reconfigured a feeder with new conductors or voltage regulating equipment our results would no longer hold.  There may be a number of relatively low cost modifications that distribution engineers could employ -- for example additional voltage regulating equipment -- that would improve feeder performance with respect to voltage excursions but increase mechanical switching events.  Optimal modification of feeders to facilitate distributed PV hosting is an important area for future research.}

\todo{14} \change{A number of other researchers have investigated the incidence of resistive losses in simulated distribution systems~\cite{quezada2006assessment, thomson2007impact, navarro2013monte, widen2010impacts}, with a very broad range of results (ranging from a large reduction in losses to an increase in losses).  Our findings capture this range, though we find that on most feeders resistive losses continue to decline up to 100 percent penetration.  Other researchers have also investigated the incidence of voltage excursions in simulation studies, and as with resistive losses our results capture the range already in the literature~\cite{thomson2007impact,navarro2013monte,widen2010impacts}.  However when we look across all our results, though some feeders have an increase in voltage excursions, most do not.  }

\change{One of the major distinguishing features of this paper is that we have investigated a very broad range of feeder types and locations with relatively high temporal and spatial resolution PV data.  This allows us to generalize our findings by investigating which factors -- in particular feeder type and location -- most strongly influence our results.  The tendency of losses to begin increasing at high penetration appears to be driven by location, but feeder type has a stronger influence on the \textit{total} reduction in resistive losses.  As one might expect, we find that percent peak load reduction depends more on location (climate) than on feeder type.  Similarly, reverse power flow depends more strongly on location than feeder type, and in general those locations with more reverse power flow are also those with more peak load reduction.   Some feeder types have little to no change in voltage magnitude deviations with increasing PV penetration, while other feeders show an increase in voltage deviations; the worst deviations occur in the same location (Sacramento).  We found that impacts on voltage regulators are small and can either increase or decrease relative to a no PV baseline, depending on feeder type (and independent of location). }

\todo{15} \change{Finally, we note that while changes in distribution planning are likely required as distributed generation increases, those changes may be required only on a small number of feeders.  This is because impacts -- both positive and negative -- are relatively small in most cases we investigated.  An important area of future research is to develop methods to identify ahead of time the locations and feeder types that will have difficulty integrating large amounts of distributed PV and to focus advanced planning on those.}

\section*{Acknowledgment}
We wish to thank John Carruthers, Donovan Currey \& Matt Heling of PG\&E; Jason Fuller and the \mbox{GridLAB-D} team at PNNL; Eric Carlson, Justin Chebahtah \& Karthik Varadarajan of SolarCity; Dan Arnold, Lloyd Cibulka, Josiah Johnston, Paul Kauzmann, Kevin Koy, James Nelson, Ciaran Roberts, Michaelangelo Tabone, Alexandra von Meier and Shuyu (Simon) Yang of UC Berkeley; Luis Fernandes and Emma Stewart of LBNL; the UCB CITRIS computing cluster for their invaluable assistance; and the editorial staff and three anonymous reviewers of this manuscript, whose comments greatly improved the clarity of the paper.  



\bibliographystyle{IEEEtran}
\bibliography{IEEEabrv,cohen_callaway_journal_physical}

\begin{thebibliography}{10}
\providecommand{\url}[1]{#1}
\csname url@samestyle\endcsname
\providecommand{\newblock}{\relax}
\providecommand{\bibinfo}[2]{#2}
\providecommand{\BIBentrySTDinterwordspacing}{\spaceskip=0pt\relax}
\providecommand{\BIBentryALTinterwordstretchfactor}{4}
\providecommand{\BIBentryALTinterwordspacing}{\spaceskip=\fontdimen2\font plus
\BIBentryALTinterwordstretchfactor\fontdimen3\font minus
  \fontdimen4\font\relax}
\providecommand{\BIBforeignlanguage}[2]{{%
\expandafter\ifx\csname l@#1\endcsname\relax
\typeout{** WARNING: IEEEtran.bst: No hyphenation pattern has been}%
\typeout{** loaded for the language `#1'. Using the pattern for}%
\typeout{** the default language instead.}%
\else
\language=\csname l@#1\endcsname
\fi
#2}}
\providecommand{\BIBdecl}{\relax}
\BIBdecl

\bibitem{katiraei2011solar}
F.~Katiraei and J.~R. Aguero, ``Solar pv integration challenges,'' \emph{Power
  and Energy Magazine, IEEE}, vol.~9, no.~3, pp. 62--71, 2011.

\bibitem{quezada2006assessment}
V.~M. Quezada, J.~R. Abbad, and T.~G.~S. Roman, ``Assessment of energy
  distribution losses for increasing penetration of distributed generation,''
  \emph{Power Systems, IEEE Transactions on}, vol.~21, no.~2, pp. 533--540,
  2006.

\bibitem{shugar1990photovoltaics}
D.~S. Shugar, ``Photovoltaics in the utility distribution system: the
  evaluation of system and distributed benefits,'' in \emph{Photovoltaic
  Specialists Conference, 1990., Conference Record of the Twenty First
  IEEE}.\hskip 1em plus 0.5em minus 0.4em\relax IEEE, 1990, pp. 836--843.

\bibitem{woyte2006voltage}
A.~Woyte, V.~Van~Thong, R.~Belmans, and J.~Nijs, ``Voltage fluctuations on
  distribution level introduced by photovoltaic systems,'' \emph{Energy
  Conversion, IEEE Transactions on}, vol.~21, no.~1, pp. 202--209, 2006.

\bibitem{thomson2007impact}
M.~Thomson and D.~Infield, ``Impact of widespread photovoltaics generation on
  distribution systems,'' \emph{Renewable Power Generation, IET}, vol.~1,
  no.~1, pp. 33--40, 2007.

\bibitem{navarro2013monte}
A.~Navarro, L.~F. Ochoa, and D.~Randles, ``Monte carlo-based assessment of pv
  impacts on real uk low voltage networks,'' in \emph{Power and Energy Society
  General Meeting (PES), 2013 IEEE}.\hskip 1em plus 0.5em minus 0.4em\relax
  IEEE, 2013, pp. 1--5.

\bibitem{widen2010impacts}
J.~Wid{\'e}n, E.~W{\"a}ckelg{\aa}rd, J.~Paatero, and P.~Lund, ``Impacts of
  distributed photovoltaics on network voltages: Stochastic simulations of
  three swedish low-voltage distribution grids,'' \emph{Electric power systems
  research}, vol.~80, no.~12, pp. 1562--1571, 2010.

\bibitem{paatero2007effects}
J.~V. Paatero and P.~D. Lund, ``Effects of large-scale photovoltaic power
  integration on electricity distribution networks,'' \emph{Renewable Energy},
  vol.~32, no.~2, pp. 216--234, 2007.

\bibitem{hoke2013steady}
A.~Hoke, R.~Butler, J.~Hambrick, and B.~Kroposki, ``Steady-state analysis of
  maximum photovoltaic penetration levels on typical distribution feeders,''
  \emph{Sustainable Energy, IEEE Transactions on}, vol.~4, no.~2, pp. 350--357,
  2013.

\bibitem{cohen2013modeling}
M.~A. Cohen and D.~S. Callaway, ``Modeling the effect of geographically diverse
  {PV} generation on {C}alifornia's distribution system,'' in \emph{Smart Grid
  Communications (SmartGridComm), 2013 IEEE International Conference on}.\hskip
  1em plus 0.5em minus 0.4em\relax IEEE, 2013, pp. 702--707.

\bibitem{schneider2008modern}
K.~P. Schneider, Y.~Chen, D.~P. Chassin, R.~G. Pratt, D.~W. Engel, and
  S.~Thompson, ``Modern grid initiative: Distribution taxonomy final report,''
  Pacific Northwest National Laboratory, Tech. Rep., 2008.

\bibitem{GLD_houseguide}
\BIBentryALTinterwordspacing
(Accessed March 2015) Pacific Northwest National Laboratory. [Online].
  Available: \url{http://gridlab-d.sourceforge.net/wiki/index.php/House}
\BIBentrySTDinterwordspacing

\bibitem{caiso2013peak}
\BIBentryALTinterwordspacing
(2013, July) California {ISO} peak load history 1998 through 2012. CAISO.
  [Online]. Available:
  \url{http://www.caiso.com/Documents/CaliforniaISOPeakLoadHistory.pdf}
\BIBentrySTDinterwordspacing

\bibitem{pnnl2012population}
\BIBentryALTinterwordspacing
(2012, Nov) Pacific Northwest National Laboratory. [Online]. Available:
  \url{http://gridlab-d.svn.sourceforge.net/viewvc/gridlab-d/Taxonomy_Feeders/PopulationScript/}
\BIBentrySTDinterwordspacing

\bibitem{schneider2010evaluation}
K.~P. Schneider, J.~Fuller, F.~Tuffner, and R.~Singh, ``Evaluation of
  conservation voltage reduction ({CVR}) on a national level,'' Pacific
  Northwest National Laboratory report, Tech. Rep., 2010.

\bibitem{shoults2012power}
R.~R. Shoults and L.~D. Swift, ``Power system loads,'' in \emph{Electric Power
  Generation, Transmission, and Distribution}, L.~L. Grigsby, Ed.\hskip 1em
  plus 0.5em minus 0.4em\relax CRC Press, 2012, ch.~20.

\bibitem{bravo2012dynamic}
R.~J. Bravo, ``Dynamic performance of residential loads,'' Master's thesis,
  California State University, Long Beach, 2012.

\bibitem{fernandes2012personal}
L.~Fernandes, personal communication, LBNL, 2012.

\bibitem{nrel2012midc}
\BIBentryALTinterwordspacing
(2012, Nov) Measurement and instrumentation data center. National Renewable
  Energy Laboratory. [Online]. Available: \url{http://www.nrel.gov/midc/}
\BIBentrySTDinterwordspacing

\bibitem{noaa2013quality}
\BIBentryALTinterwordspacing
(2013, July) Quality controlled local climatological data. NOAA. [Online].
  Available: \url{http://cdo.ncdc.noaa.gov/qclcd/QCLCD?prior=N}
\BIBentrySTDinterwordspacing

\bibitem{cohen2013taxonomy}
\BIBentryALTinterwordspacing
M.~A. Cohen. (2013, May). [Online]. Available:
  \url{http://emac.berkeley.edu/gridlabd/taxonomy_graphs/}
\BIBentrySTDinterwordspacing

\bibitem{ieee1996guide}
\emph{IEEE Guide for Loading Mineral-Oil-Immersed Transformers}, IEEE Std.
  C57.91-1995, 1996.

\bibitem{ge1972distribution}
\emph{Distribution Data Book}.\hskip 1em plus 0.5em minus 0.4em\relax General
  Electric, 1972, no. GET-1008L.

\bibitem{abb2001distribution}
\emph{Distribution Transformers}, ABB, Business Area Distribution Transformers,
  P.O. Box 8131, CH - 8050 Zurich, Switzerland, 2001.

\bibitem{GLD_powerflowguide}
\BIBentryALTinterwordspacing
(Accessed March 2015) Pacific Northwest National Laboratory. [Online].
  Available:
  \url{http://gridlab-d.sourceforge.net/wiki/index.php/Power_Flow_User_Guide}
\BIBentrySTDinterwordspacing

\bibitem{caiso2013oasis}
\BIBentryALTinterwordspacing
(2013, April) California {ISO} open access same-time information system
  ({OASIS}). CAISO. [Online]. Available: \url{http://oasis.caiso.com}
\BIBentrySTDinterwordspacing

\bibitem{carruthers2013personal}
J.~Carruthers, personal communication, PG\&E, 2013.

\bibitem{mather2012quasi}
B.~A. Mather, ``Quasi-static time-series test feeder for {PV} integration
  analysis on distribution systems,'' in \emph{Power and Energy Society General
  Meeting, 2012 IEEE}.\hskip 1em plus 0.5em minus 0.4em\relax IEEE, 2012, pp.
  1--8.

\bibitem{lave2013wavelet}
M.~Lave, J.~Kleissl, and J.~S. Stein, ``A wavelet-based variability model (wvm)
  for solar pv power plants,'' \emph{Sustainable Energy, IEEE Transactions on},
  vol.~4, no.~2, pp. 501--509, 2013.

\end{thebibliography}
\end{document}